\documentclass[12pt]{article}
\usepackage[english]{babel}
\usepackage{epsfig}
       
\begin{document}
\setcounter{page}{0}

\begin{center}
{\Huge Categorization by a Three-state Attractor Neural Network}
\\ \vspace{1.cm}
{\large D.R.C.~Dominguez 
\footnote{e-mail:david@tfdec1.fys.kuleuven.ac.be}
	and D.~Boll\'e
\footnote{e-mail: desire.bolle@fys.kuleuven.ac.be
\newline  \indent  $^{(1)}$ Also at 
Interdisciplinair Centrum voor Neurale Netwerken, K.U.Leuven}$^{(1)}$
        \\Instituut voor Theoretische Fysica,
	  K.U.\ Leuven \\
	  B-3001 Leuven, Belgium \\ }
\end{center}
\date{}
\thispagestyle{empty}

\vspace{1.cm}
\begin{abstract}
\normalsize
\noindent
The categorization properties of an attractor network of three-state
neurons which infers three-state concepts from examples are studied. The
evolution
equations governing the parallel dynamics at zero temperature for the
overlap between the state of the network and the examples, the state of the
network and the concepts as well as the neuron activity are discussed
in the limit of extreme dilution. A transition from a retrieval region
to a categorization region is found when the number of examples or their
correlations are increased. If the pattern activity is small enough,
the examples (concepts) are very well retrieved (categorized) for an
appropriate choice of the zero-activity threshold of the neurons.
\end{abstract}


\newpage

\label{sec:level1}

\section*{1. Introduction}

Some years ago a minimal modification of the Hopfield model has been
suggested
such that categorization of patterns emerges naturally from an encoding
stage structured in layers \cite{PV86}. The network is spatially homogenous
in the sense that the synaptic couplings between all neurons are of the
same type, but the patterns are hierarchically ordered, meaning that
patterns belonging to the same group are strongly correlated while patterns
sitting in distinct groups are only weakly correlated. Related models of
this type have been examined in \cite{FI}-\cite{G}. These models lead to
the appearance of stable states besides those corresponding to the original
patterns, e.g., the ancestors of the categories to which those patterns
belong.

Shortly after, a simple Hebbian rule has been proposed in order to
study the performance of a network in learning an extensive number of
ancestor patterns in such an hierarchical ordering, given that the
learning takes place with groups of a finite number of (correlated)
patterns situated on a lower level of the hierarchical tree \cite{FM89}.
In other words the problem of categorizing examples (i.e. the correlated
patterns) into classes defined by concepts (i.e. the ancestors) is studied.
It turns out that such a network looses its ability to
retrieve the examples when a critical number of them is presented during
the learning stage, but it then gets the ability to categorize the concepts
\cite{FM89}- \cite{CT95}.

This categorization property ocurring in fact through learning from
examples is thus a particular kind of generaliztion. Generalization
has been a topic of intensive research in recent years. (For recent
reviews emphasizing different aspects see \cite{SS92}-\cite{Opperk}.)

Recently, models with multi-state and analogue neurons have been introduced
in the study of categorization problems \cite{ST92,DT96}. By using
analogue neurons \cite{ST92}, less (binary) examples are needed in order to
start categorization. However, the generalization error, i.e., the Hamming
distance between the microscopic state of the network and the (binary)
concepts is larger than in the corresponding two-state model. A further
improvement is obtained by using low-activity examples, from which
(binary) full-activity concepts can be infered, even if the number of
examples is small \cite{DT96}.  This must be due to the fact that mixture
states of patterns can be inherently stable, allowing the network to
ultimately form higher activity patterns out of smaller ones, just like
happens in the retrieval regime for both highly diluted \cite{Ye89} and
fully connected three-state networks \cite{MH89}.

In this paper we extend these models by allowing the concepts themselves to
be three-state. Furthermore, we do not require full symmetry of the
retrieval overlaps of the examples. At the same time we study the
conditions characterizing the transition from the retrieval phase to the
categorization phase. These problems are considered in an asymmetrically
diluted network with a Hebbian type learning rule because, as is standard
knowledge by now, its parallel dynamics can be solved exactly \cite{DG87}.
The following main results are found. When the number of examples (per
concept) presented to the network is too small or their correlations are
too weak, it behaves as a retrieval (of examples) device.  When one of
these parameters attains a large enough value the network starts
categorizing. The latter behavior can be considerably improved by
appropriately chosing the zero-activity threshold of the neurons.
Compared with the categorization properties of the binary concept model
\cite{DT96}, we find that the categorization error is smaller and that a
greater number of concepts can be categorized.

The rest of this paper is organized as follows. In Section 2 we introduce
the model and the relevant Hamming distances as macroscopic measures for
the retrieval and categorization quality of the model. Section 3 solves
the parallel dynamics leading to evolution equations for the retrieval
overlap, the categorization overlap and the neuron activity. In Section 4
we study the retrieval and categorization phases as a function of the
the zero-activity threshold of the neurons, the number of concepts, the
number of examples per concept, their correlations and their activity.
Finally, Section 5 presents some concluding remarks.

\section*{2. The model}

Consider a network of $N$ three-state
neurons. At time $t$ and zero temperature the neurons $\{\sigma_{i,t}\}$
are updated in parallel according to the rule
\begin{eqnarray}
    \sigma_{i,t+1}&=& F_{\theta}(h_{i,t}), \quad i=1,...,N
          \label{2.sit0} \\
            h_{i,t}&=& \sum_{j(\neq i)} J_{ij}\sigma_{j,t} \, ,
         \label{2.sit}
\end{eqnarray}
where $h_{i,t}$ is the local field of neuron $i$ at time $t$. The
input-output relation $F_{\theta}$ is, in general, a monotonous function
and will later on be chosen as the three-state step-like function
\begin{equation}
     F_{\theta}(x) = \left\{ \begin{array}{ll}
                         \mbox{sign}(x) &\mbox{if $|x|>\theta $} \\
                         0         &\mbox{if $|x|<\theta $}
                     \end{array} \right.
      \label{2.Fte}
\end{equation}
where $\theta$ is the zero-activity threshold parameter of the neurons.
The synaptic couplings $J_{ij}$ are determined through the learning of
$s$ three-state examples, $\eta^{\mu\rho}_{i}\in\{0,\pm 1\}\,,
\mu=1,\ldots,p \,, \rho=1,\ldots,s $, of $p$ three-state concepts,
$\xi^{\mu}_{i}\in\{0,\pm 1\}$. The examples have zero mean and variance
$A=1/N \sum_i(\eta^{\mu\rho}_{i})^{2}$, which is a measure for their
activity. The concepts $\xi^{\mu}_{i}$ are chosen to be independent
identically
distributed random variables (i.i.d.r.v.) with mean zero and activity
equal to the activity $A$ of the examples. The following Hebbian-type
algorithm is taken
\begin{equation}
     J_{ij}={1 \over NA} \sum_{\mu=1}^{p} \sum_{\rho=1}^{s}
                      \eta^{\mu\rho}_{i} \eta^{\mu\rho}_{j}.
         \label{2.Jij}
\end{equation}
Furthermore, each set of examples $\{ \eta^{\mu\rho}_{i} \}_{\rho=1}^{s}$
at site $i=1,...,N$ is built from the concept $\xi^{\mu}_{i}$
through the following process
\begin{equation}
        \eta^{\mu\rho}_{i}= \xi^{\mu}_{i} \lambda^{\mu\rho}_{i},
              \quad \lambda^{\mu\rho}_{i} \in\{\pm 1\} \, .
       \label{2.emr}
\end{equation}
The variables $\lambda^{\mu\rho}_{i}$ are also taken to be i.i.d.r.v. with
a bias towards the value $+1$ such that they are given by the probability
distribution
\begin{equation}
        \mbox{p}(\lambda^{\mu\rho}_{i})=
           b_{+} \delta(\lambda^{\mu\rho}_{i}-1)+
           b_{-} \delta(\lambda^{\mu\rho}_{i}+1),
     \label{2.lmr}
\end{equation}
with $b_{\pm}=(1\pm b)/2$. The parameter $b$ describes the correlation
between the stored example $\eta^{\mu\rho}_i$ and its concept
$\xi^{\mu}_j$, viz.
$\langle\eta^{\mu\rho}_{i}\xi^{\mu}_{j}\rangle = bA\delta_{ij}$,
and the correlation between two different examples of the same concept
$\langle\eta^{\mu\rho}_{i}\eta^{\mu\sigma}_{j}\rangle = b^{2}A\delta_{ij}$.

At this point we remark that, on the one hand we recover the binary
categorization model \cite{FM89} by setting $A=1$ and $\theta=0$.
On the other hand, the standard three-state neuron model \cite{Ye89} is
obtained by taking the number of examples in Eq.(\ref{2.Jij}) to be $1$ and
the correlation $b=1$.

In order to measure the quality of retrieval of the examples we introduce
the Hamming distance between the stored example and the microscopic state
of the network \cite{BV93}
\begin{eqnarray}
      D^{\mu\rho}_{t} =
              {1\over N}\sum_{i}[\eta^{\mu\rho}_{i}-\sigma_{i,t}]^{2}
                     = A-2A\,m^{\mu\rho}_{N,t}+Q_{N,t} \, .
        \label{2.Dmr}
\end{eqnarray}
This defines the retrieval overlap between the microscopic state of the
network and the $\rho$th example of the $\mu$th concept
\begin{equation}
         m^{\mu\rho}_{N,t} =
              {1\over NA}\sum_{i}\eta^{\mu\rho}_{i}\sigma_{i,t}\, .
            \label{2.mmr}
\end{equation}
These are normalized order parameters within the interval $[-1,1]$,
which attain the maximal value $m^{\mu\rho}_{ N}=1$ whenever
$\sigma_{i}= \eta^{\mu\rho}_{i}$ (recall Eq.(\ref{2.emr})).
Furthermore, also the neuron activity is introduced
\begin{equation}
         Q_{N,t} = {1\over N}\sum_{i}|\sigma_{i,t}|^{2}\, .
           \label{2.QNt}
\end{equation}

The task of categorization is successful when the distance between the
microscopic state of the network and the concept $\xi_i^{\mu}$, defined as
\begin{eqnarray}
          E^{\mu}_{t} =
            {1\over N}\sum_{i}[\xi^{\mu}_{i}-\sigma_{i,t}]^{2}
                   = A-2AM^{\mu}_{N,t}+Q_{N,t}
        \label{2.Emt}
\end{eqnarray}
becomes small after some time $t$. As explained in the introduction the
quantity $E^{\mu}_{t}$ can be considered as the generalization error in
this context. (We refer to the refs.~(\cite{SS92}-\cite{Opperk}) for a
comparison with definitions of the generalization error in related
contexts of learning from examples.) In obtaining the second equality of
Eq.~\ref{2.Emt} we have used the fact that the activity of the concepts
and the examples are taken to be equal. Furthermore, the overlap between
the microscopic state of the network and the concept is defined as
\begin{equation}
      M^{\mu}_{N,t} = {1\over NA}\sum_{i}\xi^{\mu}_{i}\sigma_{i,t}\, .
      \label{2.MmN}
\end{equation}

We now want to consider an extremely diluted asymmetric version of this
model in which each neuron is connected, on average, with $C$ other
neurons through the synaptic couplings (see expression (\ref{2.Jij}))
\begin{eqnarray}
     J_{ij}(C) = C_{ij} {N\over C} J_{ij}
             ={C_{ij}\over CA} \sum_{\mu=1}^{p} \sum_{\rho=1}^{s}
                      \eta^{\mu\rho}_{i} \eta^{\mu\rho}_{j}.
     \label{2.JijC}
\end{eqnarray}
Here, the $C_{ij}\in\{0,1\}$ are i.i.d.r.v. with probability
$\mbox{Pr}\{C_{ij}=1\}=C/N, C>0$. These $C_{ij}$ are highly asymmetric
such that in the limit of extreme dilution, $C << \log N$, the
architecture of the network gets the structure of a directed tree and
the neurons are uncorrelated for almost all sites $i$. This allows an
exact solution of the parallell dynamics \cite{DG87}.

In the next section we discuss this solution by writing down the evolution
equations for the retrieval overlap, the neuron activity and the
categorization overlap.

\section*{3. The diluted dynamics}

Because we are interested in both the retrieval and categorization
properties of the network, we take an initial configuration correlated with
only one concept meaning that only the retrieval overlaps for the $s$
examples of that given concept, say the first one, are macroscopic,
i.e., of order $O(1)$ in the thermodynamic limit $N \rightarrow \infty$. In
order to study the retrieval of a particular example we single out the
component $\rho=1$. We furthermore assume that
all other components are the same i.e., $m^{1\rho }_{N,t}= m^{1s}_{N,t}$
for all  $\rho>1$.
This property of the examples is called quasi-symmetry. It extends former
work (e.g., \cite{DT96}) where full symmetry of the examples has been
assumed.

The dynamics of this model is then studied following standard methods
involving a signal-to-noise analysis (see, e.g., \cite{DG87}, \cite{BV93},
\cite{BSVZ94}). At this point we recall that it is justified to first
dilute the system by taking the limit $N \rightarrow \infty$ and second, in
the diluted system, to apply the law of large numbers (LLN) and the central
limit theorem (CLT) by taking the limit $C \rightarrow \infty$. Furthermore
the retrieval overlaps have to be considered over the diluted structure and
the loading $\alpha$ is defined by $p=\alpha C$.
Finally, we know that because of the extremely diluted structure of the
network the equations derived for the first time step are valid for
any time step.

Splitting the local field (\ref{2.sit}) into a signal and noise part gives
\begin{equation}
  h_{i,t} = \eta^{11}_{i} m^{11}_{C,t} +
       \sum_{\rho > 1}^{s} \eta^{1\rho}_{i} m^{1\rho}_{C,t} +
       \sum_{\mu > 1}^{p}\sum_{\rho}^{s} \eta^{\mu\rho}_{i}
       \sum_{j \neq i}^N \frac{C_{ij}}{CA} \eta^{\mu\rho}_{j}\sigma_{j,t}
         \label{3.hit}
\end{equation}
with
\begin{equation}
   m^{1\rho}_{C,t}= \sum_{j(\neq i)}^{N} {C_{ij}\over CA}
                \eta^{1\rho}_{j} \sigma_{j,t} \, .
         \label{3.mC}
\end{equation}
In the thermodynamic limit we then obtain in a standard way (\cite{DT96},
\cite{DG87}, \cite{BV93}, \cite{BSVZ94})
\begin{eqnarray}
    m^{11}_{t+1} &=& \langle\langle\langle
         \lambda^{11}  F_{\theta}({\tilde h}_{t})
          \rangle_{\lambda^{11}} \rangle_{x_{s}} \rangle_{\omega_{t}}
               \label{3.m110} \\
    m^{1s}_{t+1} &=& \langle\langle\langle
           x_{s} F_{\theta}({\tilde h}_{t})
          \rangle_{\lambda^{11}} \rangle_{x_{s}} \rangle_{\omega_{t}}
               \label{3.m11} \\
    M^{1}_{t+1} &=& \langle\langle\langle F_{\theta}({\tilde h}_{t})
          \rangle_{\lambda^{11}}\rangle_{x_{s}}\rangle_{\omega_{t}}
               \label{3.Mt+} \\
    Q_{t+1} &=& A \langle\langle\langle [F_{\theta}({\tilde h}_{t})]^{2}
           \rangle_{\lambda^{11}}\rangle _{x_{s}}\rangle_{\omega_{t}}
     + (1-A)\langle [F_{\theta}(\omega_{t})]^{2}\rangle _{\omega_{t}} \,.
               \label{3.Qt+}
\end{eqnarray}
with
\begin{eqnarray}
    {\tilde h}_{t} &\doteq&
       \lambda_i^{11} m^{11}_{t} + (s-1) x_{s} m^{1s}_{t}+ \omega_{t}
          \label{3.Lte0}  \\
     \omega_{t} &=&
         [ \alpha r Q_{t}]^{1/2} {\cal N}(0,1) \, .
         \label{3.Lte}
\end{eqnarray}
Here $\doteq$ indicates that this relation is valid in distribution,
 $x_{s}= {1\over s-1} \sum_{\rho>1}^{s} \lambda^{1\rho}$, $r=
s(1+(s-1)b^{4})$ and
the quantity ${\cal N}(0,1)$ is a Gaussian random variable with mean zero
and variance unity. Furthermore, we have averaged already over $\xi^{1}$.
The brackets denote the further averages
over both $\lambda^{11}$, $x_{s}$ and over $\omega_{t}$. We
recall that the average over $\lambda^{11}$ has to be done according to the
distribution (\ref{2.lmr}).

The first term in the expression (\ref{3.Lte0}) is the signal coming
from the first example of the first concept, while the second term,
recalling the assumed quasi-symmetry of the examples, represents the signal
of the other examples of the first concept. It has a strength factor
$x_{s}$. The third term
is the noise caused by the examples of the $(p-1)$ residual non-condensed
concepts.

These Eqs.(\ref{3.m110})-(\ref{3.Qt+}) give a complete description of the
dynamics for the retrieval of examples and the categorization of the
concepts by the network we are considering for a general monotonous
input-output function $F_{\theta}$. Chosing $F_{\theta}$ to be the
three-state function (\ref{2.Fte}) we obtain the following explicit forms
for the dynamics
\begin{eqnarray}
   m^{11}_{t+1}&=&\sum_{j=0}^{s-1}p_{b}(j)
     \{b_{+}[\Psi^{+}_t(\Omega_{+})+\Psi^{-}_t(\Omega_{+})]
      -b_{-}[\Psi^{+}_t(\Omega_{-})+\Psi^{-}_t(\Omega_{-})]\}
          \label{3.mmM0}  \\
   m^{1s}_{t+1}&=&\sum_{j=0}^{s-1}p_{b}(j){2j-s+1\over s-1}
     \{b_{+}[\Psi^{+}_t(\Omega_{+})+\Psi^{-}_t(\Omega_{+})]
      + b_{-}[\Psi^{+}_t(\Omega_{-})+\Psi^{-}_t(\Omega_{-})]\}
          \nonumber\\    \\
    M^{1}_{t+1}&=&\sum_{j=0}^{s-1}p_{b}(j)s
       \{b_{+}[\Psi^{+}_t(\Omega_{+})+\Psi^{-}_t(\Omega_{+})]
       + b_{-}[\Psi^{+}_t(\Omega_{-})+\Psi^{-}_t(\Omega_{-})]\}
               \\
    Q_{t+1}&=& 1-A\sum_{j=0}^{s-1}p_{b}(j)
       \{b_{+}[\Psi^{+}_t(\Omega_{+})-\Psi^{-}_t(\Omega_{+})]
       + b_{-}[\Psi^{+}_t(\Omega_{-})-\Psi^{-}_t(\Omega_{-})]\}
          \nonumber\\
       && -2(1-A)[\Psi^{+}_t(0)].
     \label{3.mmM}
\end{eqnarray}
Here
\begin{equation}
     p_{b}(j)=\left(\begin{array}{c} s-1\\j \end{array}\right)
               b_{+}^{j}b_{-}^{s-1-j},
\label{3.pBj}
\end{equation}
and
\begin{equation}
    \Psi^{\pm}_t(\Omega_{\pm}) =
       \mbox{erf}({\Omega_{\pm}\pm\theta\over\sqrt{\alpha r Q_{t}}}),
       \quad \Omega_{\pm}=(2j-s+1)m^{1s}_{t}\pm m^{11}_{t},
     \label{3.erf}
\end{equation}
where $\mbox{erf}(x)=\int_{0}^{x}dz e^{-z^{2}/2}/\sqrt{2\pi}$.
In the case that we have many examples per concept we use for $x_s$ defined
before the
Gaussian approximation $x_{s}\doteq {b} + z_{s}\sqrt{{1-b^{2}\over s-1}}$
with $z_{s}= {\cal N}(0,1)$ independent of $\omega_t$.

\section*{4. Retrieval and categorization}

We now discuss the structure of the retrieval and categorization dynamics
which can be extracted by numerical solution of the fixed-point equations
given by (\ref{3.mmM0})-(\ref{3.mmM}) (by leaving out the $t$-dependence).

Besides the zero solution $Z$ determined by $m^{11}=M^1=Q=0$ it is necesary
to distinguish among the following different types of solution: the
retrieval solutions $R$ defined by $m^{11}>M^1>0$ and $Q>0$, the
categorization solutions $G$ defined by $0<m^{11}<M^1$ and $Q>0$ and the
self-sustained activity \cite{BE93} solutions $S$ with $Q>0$ but
$m^{11}=M^1=0$. For
the retrieval respectively categorization solution we impose the further
condition $D<0.1$ respectively $E<0.1$. Hereby the numbers $0.1$ are
somewhat arbitrarily chosen but the idea is to guarantee a minimal
retrieval respectively generalization quality of the network.

Since there are many parameters to be considered in the discussion of
the numerical results we only show in  Figs.~1-6 the properties of the
network we believe to be typical and important.

Figure 1 shows the Hamming distance $D= D^{11}_{\infty}$, the
generalization error $E= E^{1}_{\infty}$ and the neuron activity
$Q= Q_{\infty}$ as a function of the correlation $b$. The other parameters
are chosen as follows: the number of examples $s=5$, the loading rate
$\alpha=0.01$, the activity $A$ takes the values $1$ (binary) in the
upper part of the figure and $0.3$ (three-state) in the lower part, and the
zero-activity threshold $\theta$ is $0$ (binary) in the left part of the
figure and $0.5$ (three-state) in the right part.
For binary patterns it is seen that the use of three-state neurons does
not affect the overall behaviour of the network. However, for three-state
patterns both the retrieval and the categorization abilities are improved.
The transition from a $R$ phase to a $G$ phase is clearly present in all
data. For a critical value of the correlation, $b_{c}$, there is a crossing
between the $D$ and $E$ lines.
Here we remark that in the case of three-state patterns $(A=0.3)$ and
binary neurons $(\theta=0)$ the conditions for good retrieval and
categorization behaviour of the full patterns are, of course, not
satisfied. But as the curves indicate, e.g., one finds the best possible
retrieval of the active sites ($D=0.7$ for $b<b_c$). Furthermore, we also
note
the existence of a plateau for $D$ in the case of three-state patterns and
three-state neurons, where the Hamming distance is not small but it still
satisfies $D<E$. Finally, in all cases there exists a minimal value for $E$
meaning that the categorization is optimal for the corresponding network
parameters. This does not always happen for $b=1$, the reason being that
although the neuron activity $Q$ becomes high for large $b$, the pattern
activity $A$ may be so small that $\sigma_i$ can not match $\xi_i$.
This is in agreement with Eq.(\ref{2.Emt}).

This behaviour is further illustrated in typical $(\theta,b)$ and $(A,b)$
phase diagrams, Fig.~2 respectively Fig.~3, where the different phases
corresponding to the solutions of the fixed-point equations described
before are shown. We have divided the $R$-phase in two regions, one of
them ($R_{+}$) refering to the region where $D$ is almost zero, the other
one ($R_{-}$) indicating the region where $D$ has already jumped to the
plateau seen in Fig.~1. The thin dashed line $G_{opt}$ in the phase
diagrams
describes the optimal categorization. We remark the existence of two
$S$-phases in Fig.~3, the first one separating the $R$- and $G$-phase,
the second one occuring for large correlations $b$. The first indicates
a region where $b$ is already too large to have retrieval but still too
small to allow categorization. The second describes the high neuron
activity region for large $b$ mentioned above.

In Fig.~4 we plot the behaviour of the network as a function of the
zero-activity threshold $\theta$ for $A=0.01$ and $s=80$ examples per
concept. On the left we show both the retrieval overlap $m$ and the
overlap with a concept $M$ for $\alpha=0.02$ and small correlations
$b=0.1$ such that a $R$-phase exists.
For an appropriate choice of the threshold $m\approx 1$ while $M$ becomes
small. Thereby we note that although the concept storage
seems to be rather small ($\alpha=p/C=0.02$), the example storage
 is large for correlated patterns ($\alpha_{s}= s\alpha=1.6$).
On the right we display $M$ for several values of $\alpha$ and large
correlations ($b=0.5$) such that we are in the $G$-phase. For increasing
values of the threshold until $\theta=\theta_{opt}(\alpha)$ the overlap
with a concept becomes larger indicating that the categorization ability
inproves. For $\theta>\theta_{opt}(\alpha)$ this categorization
ability slowly decreases and at $\theta= \theta_{Z}(\alpha)$, the overlap
$M$ falls abruptly to zero. This illustrates that for a very carefully
tuned $\theta$, $M\approx 1$, implying that categorization stays succesful
even for a concept loading larger than $\alpha=3$. So
compared with the categorization properties of the binary concept model
(see Fig.~2 of \cite{DT96}), we find that by using three-state concepts the
categorization error is smaller and that a greater number of concepts can
be categorized.

Next, an $(\alpha,\theta)$ phase diagram is shown in Fig.~5. We take
$A=0.01$, $b=0.5$ and a large number of examples $s=80$ such that no
$R$-phase appears. Categorization is then possible within the full
line boundaries. Again the line $G_{opt}$ indicates the values of
$\theta(\alpha)$ for which $M$ is maximal, thus $E$ is minimal and hence
the categorization is optimal.
In the region between the full line and the (thick) dashed line, $M$ is
also of order $1$, however $Q$ is of the same order such that
the condition for good categorization, $E<0.1$, is not satisfied.
Above the dashed-dotted line an $S$-phase does exist. Phase coexistence is
possible in some regions and precisely what phase is attained depends on
the initial conditions $m^{11}_0, m^{1s}_0$ and $Q_0$.

Finally, in Fig.~6 we show $m$ and $M$ as a function of $\theta$ for an
analogue input-output relation $F_{\theta}=\tanh(x/\theta)$. For
comparison, the same network parameters are used as in Fig.~4 for the
three-state case and $\alpha$ again takes several values. Concerning
retrieval of the examples a similar behaviour is found as in Fig.~4
with a slightly smaller $m$. Concerning categorization, however,
although an analogous non-monotonous behaviour in $\theta$ is seen, $M$
does not come close to $1$ for larger $\alpha$. It demonstrates that the
gain parameter of a continuous input-output relation does not play the
role the zero-activity threshold does for the three-state case. The reason
is that in the three-state case this threshold switches of the neurons
whose field $h_{i}$ is not large enough such that the $\sigma_{i}$ can
match the three-state patterns $\xi_{i}$.

\section*{5. Concluding remarks}

We have studied the retrieval and categorization properties of an extremely
diluted three-state neural network through the solution of its parallel
dynamics. In comparison with existing models in the literature the
concepts are allowed to be three-state and the retrieval overlaps
between the examples and the microscopic state of the network are not
assumed to be fully symmetric. We find that
the important parameters governing the transition from the retrieval to the
categorization phase are the number of examples per concept and their
correlations. By chosing appropriately the zero-activity threshold of the
neurons categorization can be considerably improved. In particular, in
comparison with models for binary concepts the categorization error is
smaller and a much greater number of concepts can be categorized.

\section*{Acknowledgments}

We thank S.Amari, E.Koroutcheva, N.Parga and W.K.Theumann for useful
discussions. This work has been supported in part by Cnpq/Brazil, the
Universidad Autonoma de Madrid, and the Research Fund of the K.U.Leuven
(grant OT/94/9). One of us (D.B.) is indebted to the Fund for Scientific
Research - Flanders (Belgium) for financial support.




\begin{figure}[t]
\epsfxsize=15.cm
\epsfysize=11cm
\centerline{{\epsfbox[1 1 650 400]{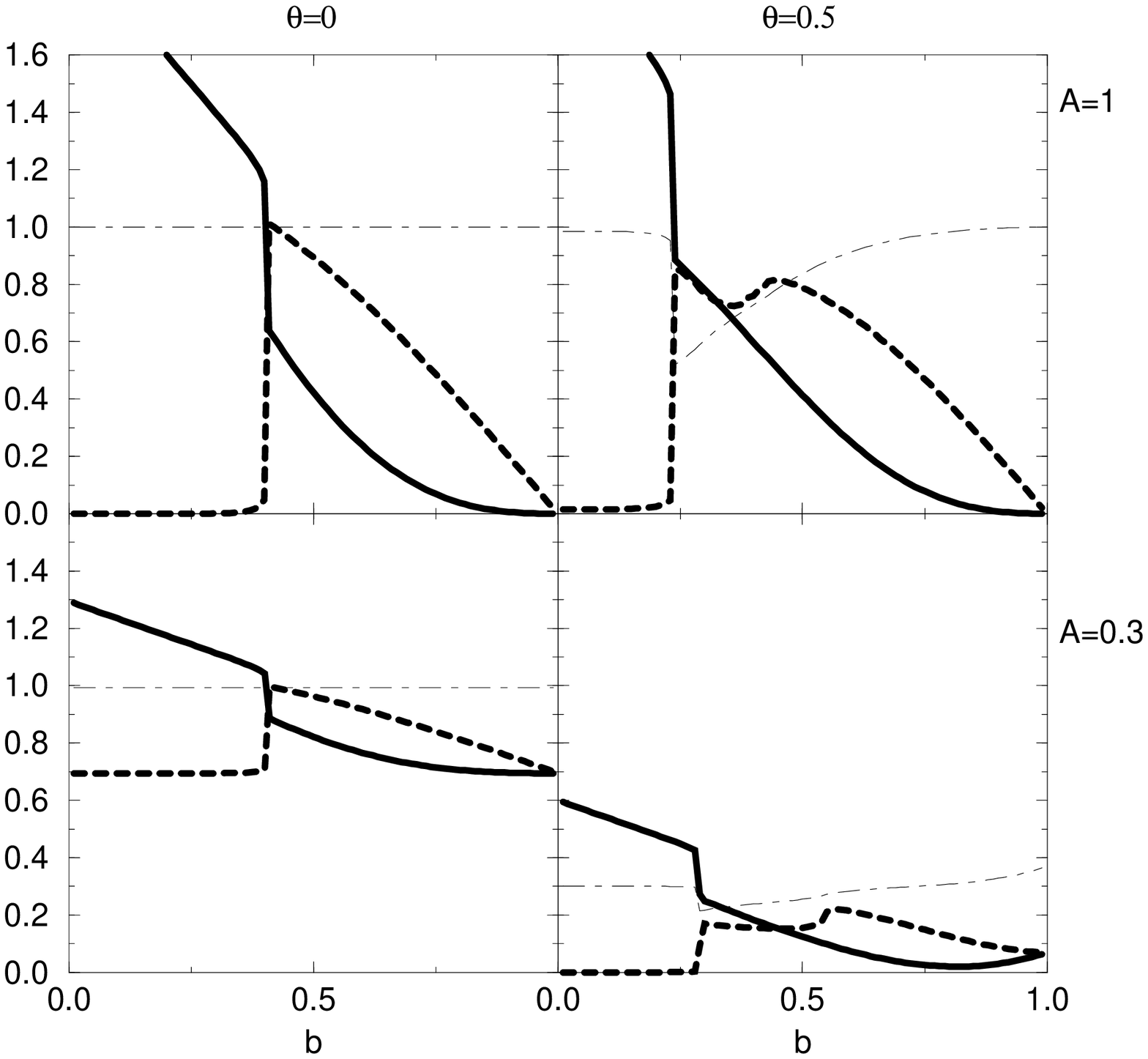}}}
\caption{ {\em The Hamming distance $D$ (dashed line), the categorization
error $E$ (full line) and the neuron activity $Q$ (thin dashed-dotted line)
as a function of the correlation $b$. The other network parameters
are chosen as follows: the number of examples $s=5$, the loading rate
$\alpha=0.01$, the activity $A=1$ in the upper part and
$A=0.3$ in the lower part, and the zero-activity threshold $\theta=0$
in the left part and $\theta= 0.5$ in the right part.} }
\label{0Db}
\end{figure}

\begin{figure}[t]
\epsfxsize=15.cm
\epsfysize=11cm
\centerline{{\epsfbox[10 10 650 480]{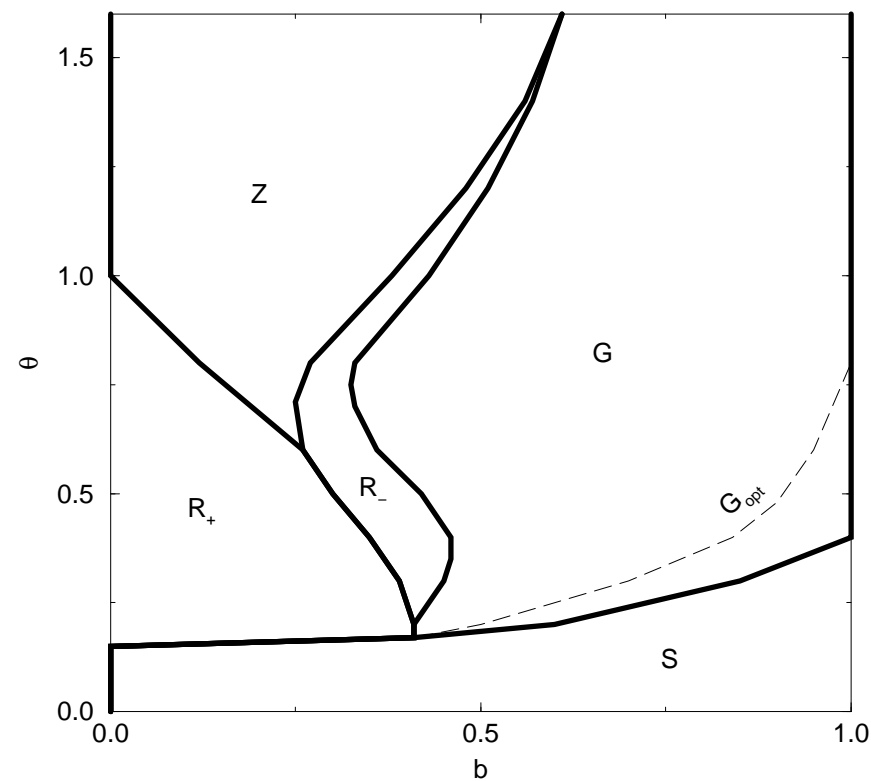}}}
\caption{ {\em The $(\theta,b)$ phase diagram with $A=0.1, s=5$ and
$\alpha=0.01$. The following phases occur: the retrieval phases $R_+$
and $R_-$, the categorization phase $G$, the self-sustained activity phase
$S$ and the phase $Z$ corresponding with the fixed-point zero. The
thin dashed line $G_{opt}$ indicates optimal categorization.} }
\label{1tb}
\end{figure}

\begin{figure}[t]
\epsfxsize=15.cm
\epsfysize=11cm
\centerline{{\epsfbox[10 10 650 480]{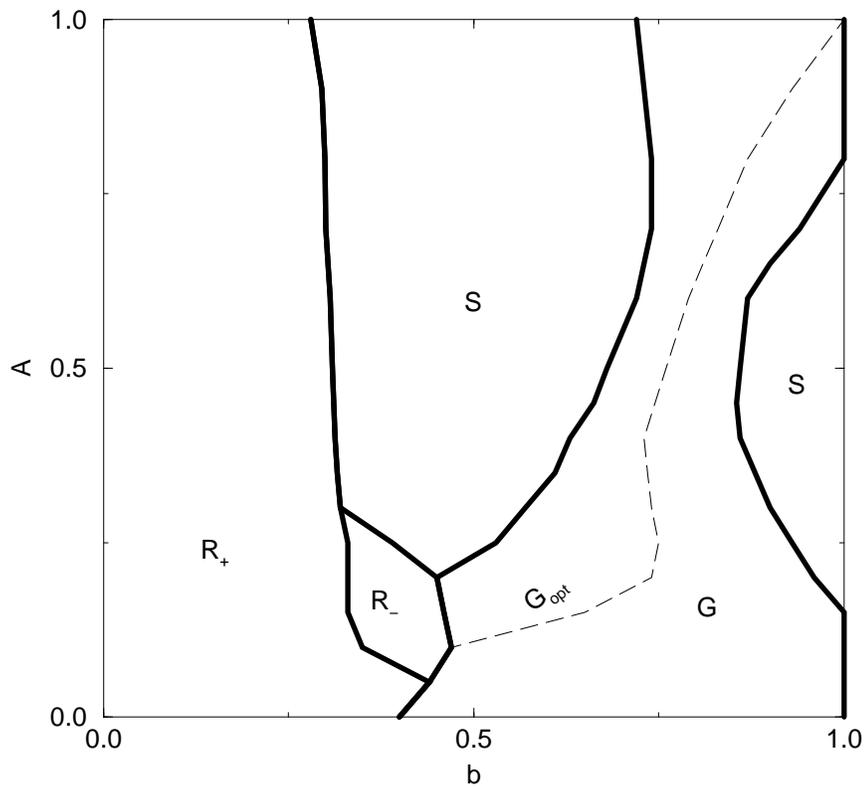}}}
\caption{ {\em The $(A,b)$ phase diagram with $\theta=0.5, s=5$ and
$\alpha=0.01$. The lines are as in Fig.~2.} }
\label{2Ab}
\end{figure}

\begin{figure}[t]
\epsfxsize=15.cm
\epsfysize=11cm
\centerline{{\epsfbox[10 10 650 400]{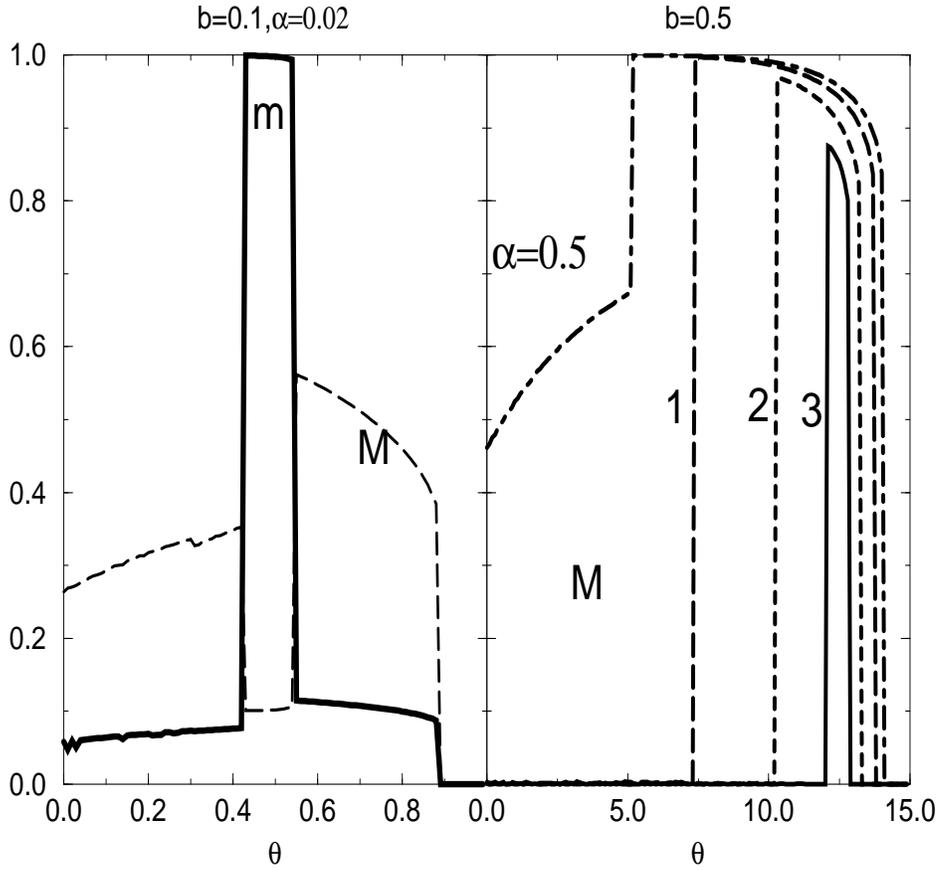}}}
\caption{ {\em $Left:$ The overlaps $m$ (full line) and $M$ (dashed line)
as a function of $\theta$ for $A=0.01$, $s=80$, $b=0.1$ and $\alpha=0.02$.
$Right:$ The overlap $M$ as a function of $\theta$ for $A=0.01$, $s=80$,
$b=0.5$ and $\alpha=0.5$ (dashed-dotted line), $\alpha=1$ (dashed line),
$\alpha=2$ (dotted line) and $\alpha=3$ (full line).} }
\label{3mt}
\end{figure}

\begin{figure}[t]
\epsfxsize=15.cm
\epsfysize=11cm
\centerline{{\epsfbox[10 10 650 400]{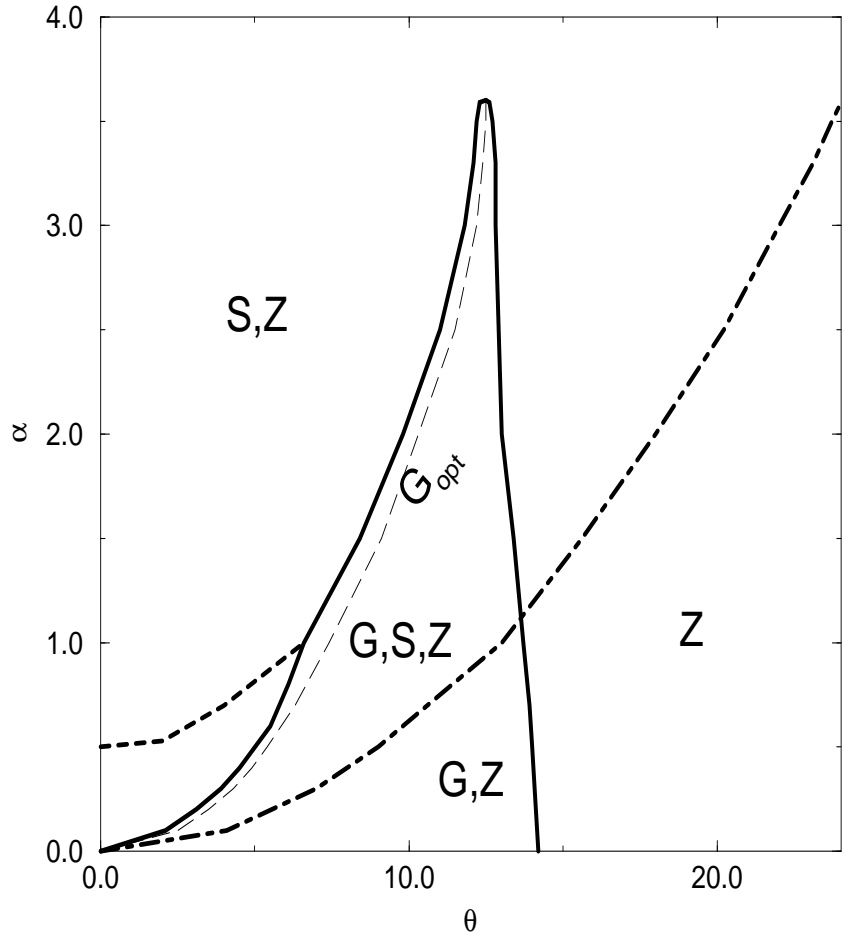}}}
\caption{ {\em
The $(\theta,\alpha)$ phase diagram with $A=0.01$, $s=80$ and
$b=0.5$. Inside the full line the $G$-phase exists. 
Between the (thick) dashed line and the full line the condition $E<0.1$ 
is not satisfied. 
Below the dashed-dotted line there exist no $S$-phase. 
The line $G_{opt}$ is as in Fig.~2.}  }
\label{4a,t}
\end{figure}

\begin{figure}[t]
\epsfxsize=15.cm
\epsfysize=11.cm
\centerline{{\epsfbox[10 10 650 400]{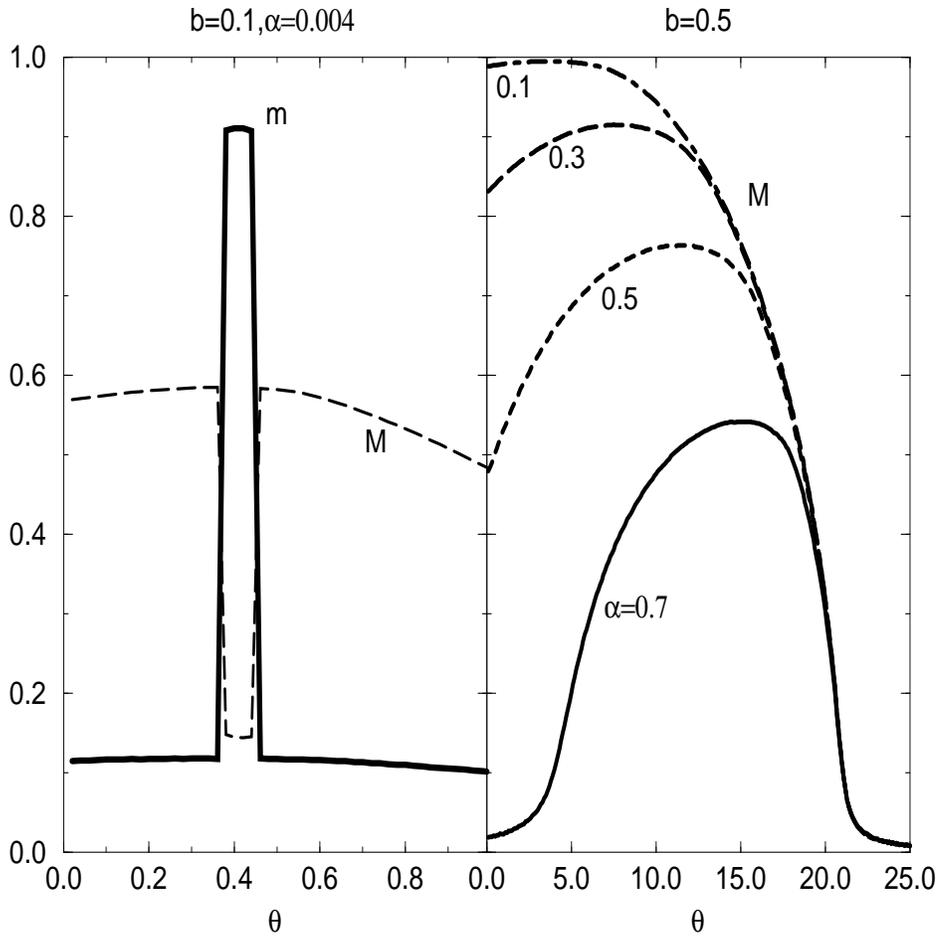}}}
\caption{{\em $Left:$ The overlaps $m$ (full line) and $M$ (dashed line) as
a function of $\theta$ for analogue neurons with 
$A=0.01$, $s=80$, $b=0.5$ and $\alpha=0.004$. 
$Right:$ The overlap $M$ as a function of $\theta$ for 
$A=0.01$, $s=80$, $b=0.5$ and 
$\alpha=0.1$ (dashed-dotted line), $\alpha=0.3$ (dashed line), 
$\alpha=0.5$ (dotted line) and $\alpha=0.7$ (full line).} }
\label{5M,t}
\end{figure}

\end{document}